\documentclass{ifacconf}

\usepackage{graphicx, color}      
\usepackage{natbib}        
\usepackage{amsmath,amssymb,amsfonts}
\usepackage{bbm}
\usepackage{amsfonts}
\usepackage{booktabs}
\usepackage{dsfont}
\usepackage{siunitx}
\usepackage{balance}
\usepackage{soul}
\usepackage{mathtools}
\usepackage{float}
\usepackage{tikz}
\usetikzlibrary{arrows.meta,positioning}
\usepackage{algorithm}
\usepackage{algpseudocode}
\algrenewcommand\algorithmicindent{1.0em}

\usepackage{tikz}
\usetikzlibrary{arrows,automata}
\usetikzlibrary{arrows,shapes,backgrounds,calc,positioning,patterns}
\usetikzlibrary{calc,arrows,shapes,backgrounds,calc,positioning,patterns,decorations.pathmorphing,decorations.markings,mindmap,trees}
\tikzstyle{block} = [draw, rectangle, minimum height=2em, minimum
width=4em] \tikzstyle{sum} = [draw, fill=blue!20, circle, node
distance=1cm] \tikzstyle{input} = [coordinate] \tikzstyle{output} =
[coordinate] \tikzstyle{pinstyle} = [pin edge={to-,thin,black}]
\usepackage[american,cute inductors,smartlabels]{circuitikz}
\usetikzlibrary{arrows,automata}
\usepackage[american,cuteinductors,smartlabels]{circuitikz}
\usetikzlibrary{calc}
\ctikzset{bipoles/thickness=1} \ctikzset{bipoles/length=0.8cm}
\ctikzset{bipoles/diode/height=.375}
\ctikzset{bipoles/diode/width=.3}
\ctikzset{tripoles/thyristor/height=.8}
\ctikzset{tripoles/thyristor/width=1}
\ctikzset{bipoles/vsourceam/height/.initial=.7}
\ctikzset{bipoles/vsourceam/width/.initial=.7} \tikzstyle{every
node}=[font=\small] \tikzstyle{every path}=[line width=0.8pt,line
cap=round,line join=round]

\begin{document}
\begin{frontmatter}

\title{Distributionally Robust Multi-Agent Reinforcement Learning for Intelligent Traffic Control}


\author[CN]{Shuwei Pei}
\author[CN]{Joran Borger}
\author[TK]{Arda Kosay}
\author[TK]{Muhammed O. Sayin}
\author[CN]{Saeed Ahmed}

\address[CN]{Jan C. Willems Center for Systems and Control, ENTEG, Faculty of Science and Engineering, University of Groningen, 9747 AG Groningen, the Netherlands (e-mails: s.pei@rug.nl, j.borger.3@student.rug.nl, s.ahmed@rug.nl)}

\address[TK]{Department of Electrical, Electronics Engineering, Bilkent University, TR-06800 Ankara, Turkey (e-mail: arda.kosay@bilkent.edu.tr, sayin@ee.bilkent.edu.tr)}

\begin{abstract}
Learning-based traffic signal control is typically optimized for average performance under a few nominal demand patterns, which can result in poor behavior under atypical traffic conditions. To address this, we develop a distributionally robust multi-agent reinforcement learning framework for signal control on a  $3\times 3$ urban grid calibrated from a contiguous $3\times3$ subarea of central Athens covered by the pNEUMA trajectory dataset~\citep{BarmpounakisGeroliminis2020}. Our approach proceeds in three stages. First, we train a baseline multi-agent RL controller in which each intersection is governed by a proximal policy optimization agent with discrete signal phases, using a centralized training, decentralized execution paradigm. Second, to capture demand uncertainty, we construct eight heterogeneous origin–destination based traffic scenarios—one directly derived from pNEUMA and seven synthetically generated—to span a wide range of spatial and temporal demand patterns. Over this scenario set, we train a contextual-bandit worst-case estimator that assigns mixture weights to estimate adversarial demand distributions conditioned on context. Finally, without modifying the controller architecture, we fine-tune the baseline multi-agent reinforcement learning agents under these estimated worst-case mixtures to obtain a distributionally robust multi-agent reinforcement learning controller. Across all eight scenarios, as well as on an unseen validation network based on the Sioux Falls configuration, the distributionally robust multi-agent reinforcement learning controller consistently reduces horizon-averaged queues and increases average speeds relative to the baseline, achieving up to 51\% shorter queues and 38\% higher speeds on the worst-performing scenarios.
\end{abstract}



\end{frontmatter}

\section{Introduction}

Urban intersections are major bottlenecks in road networks and contribute disproportionately to delay, fuel consumption, and emissions. Congested signalized junctions cause substantial economic losses through wasted time and fuel~\citep{faheem2024impact} and worsen urban air quality and public health~\citep{cohen2005global}. Rapid urbanization further increases pressure on urban transport systems~\citep{dijkstra2021applying}. Designing signal control strategies that remain effective under strongly time-varying and uncertain traffic conditions is therefore a central challenge for sustainable urban mobility and it is the main focus of this paper.

Most existing signal-control systems are based on \emph{fixed-time}, \emph{actuated}, or rule-based \emph{adaptive} control. Classical coordination schemes such as SCOOT and SCATS optimize cycle length, green splits, and offsets from historical volumes and limited detector data~\citep{Hunt1982SCOOT,Luk1984SCATSCOOT}. These methods can provide efficient progression under their design conditions, but performance degrades when actual demand deviates from the assumed patterns, for example, during incidents or atypical flows. Fully actuated and adaptive schemes improve local responsiveness, yet remain largely myopic and can struggle in dense networks where spillback, blocking, and safety constraints interact in complex ways~\citep{stevanovic2010atcs}.

Reinforcement learning (RL) offers a data-driven alternative that can adapt signal timing directly from interaction with traffic~\citep{sutton1998reinforcement}. For single intersections, deep RL controllers with compact lane- or image-based encodings, discrete phase-switching actions, and delay-oriented rewards, have demonstrated improvements over fixed-time baselines in simulation~\citep{huang2023reinforcement}. To handle networks, multi-agent reinforcement learning (MARL) assigns an agent to each intersection and coordinates them via local observations and limited neighborhood context, achieving promising results on grid and arterial networks~\citep{zhang2023learning}. These methods, however, are typically trained on a small set of nominal demand patterns and optimized for expected return, with evaluation focused on average performance.

In practice, operators care strongly about \emph{worst-case} behaviour across diverse traffic demand scenarios (e.g., peak-hours or disrupted conditions), not only average delay. Standard RL objectives provide no guarantees on tail performance, motivating robust and distributionally robust formulations that bias learning toward hard cases by upweighting poorly performing scenario groups~\citep{sagawa2020gdro}. To address this, in this paper, we adopt a contextual-bandit worst-case estimator (CB-WCE)~\citep{liu2025distributionallyrobustmultiagentreinforcement}, which adaptively reweights traffic scenarios during training and thus steers a standard MARL controller trained with proximal policy optimization (PPO) towards improved worst-case performance, while keeping the policy architecture and environment unchanged. While robust and distributionally robust RL methods have been developed and tested mainly on abstract benchmarks with simplified dynamics and unconstrained action spaces~\citep{sagawa2020gdro,hashimoto2018fairness}, it is still unclear how well these techniques transfer to network-level traffic signal control, where demand is highly variable and signal phases must satisfy strict safety constraints.

\begin{figure}[t!]
    \centering
    \includegraphics[width=1.0\linewidth]{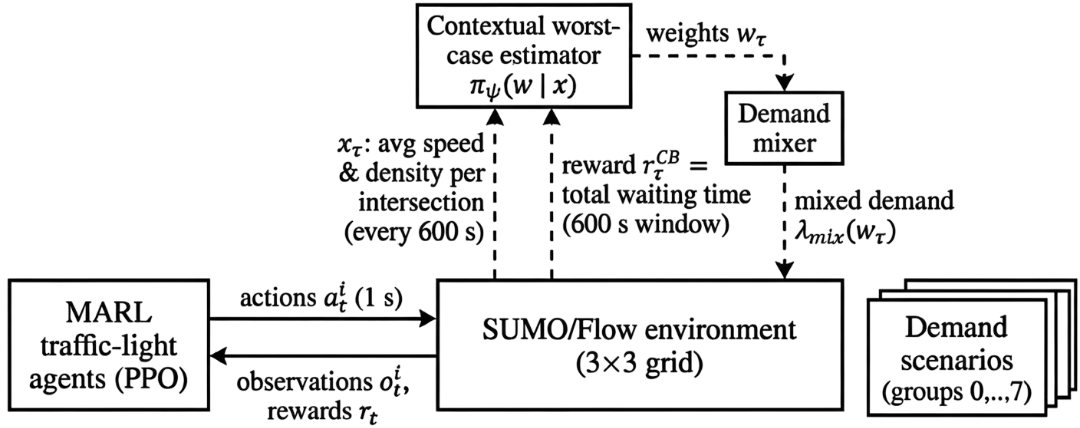}
    \caption{Schematic of DR-MARL controller for intelligent intersection management.}
    \label{fig:overview}
\end{figure}

\textbf{Contribution:}
As shown in Fig.~\ref{fig:overview}, starting from a standard MARL controller for signalized networks, we employ a CB-WCE that reweights traffic scenarios during training, yielding a distributionally robust MARL (DR-MARL) variant without changing the underlying policy architecture or environment. Concretely, our contributions are:
\begin{enumerate}
  \item We instantiate a MARL controller for a signalized \(3\times 3\) grid in a centralized training decentralized execution (CTDE) framework as our baseline MARL controller. It uses compact lane-based observations and discrete phase actions in a high-fidelity SUMO simulation environment.
  \item We propose a CB-WCE trained by OD-based demand patterns, enabling the generation of more variable and adversarial traffic scenarios for robust controller training.
  \item  We retrain the baseline MARL controller with a worst-case estimator, obtaining a DR-MARL policy, and compare its average and worst-case network performance (delay, queues, throughput) against the original MARL across various traffic scenarios.
 
\end{enumerate}

\textbf{Organization:}
Section~2 formulates the multi-agent traffic-signal control problem on the calibrated \(3\times 3\) grid and introduces the baseline MARL controller. Section~3 presents the scenario-based robustness objective, the CB-WCE, and the DR-MARL retraining procedure. Section~4 describes the experimental setup, including the network, traffic demand construction, and training protocol. Section~5 discusses numerical results, and Section~6 concludes the findings and outlines future work.


\section{Problem Setting and MARL Baseline}
\label{sec:problem}


\subsection{Traffic signal control as a multi-agent RL problem}
\label{subsec:marl-setting}

We model network-wide traffic signal control on a $3\times3$ urban grid as a MARL problem. The grid geometry (link lengths, intersection spacing) is calibrated from a contiguous $3\times3$ subarea of central Athens covered by the pNEUMA trajectory dataset~\citep{BarmpounakisGeroliminis2020}, yielding realistic block lengths and intersection density. Traffic dynamics are simulated in \textsc{SUMO}~\citep{Krajzewicz2012SUMO} and interfaced via \textsc{FLOW}~\citep{Wu2021Flow} as a Markov decision process (MDP) for RL training. Details of the simulator stack and network construction are given in Section~\ref{sec:setup}.

We model each signalized intersection as a single learning agent, with all agents treated as homogeneous and sharing a common policy that is updated together.
Let $\mathcal{I} \coloneqq \{1,\dots,9\}$ index the junctions, and let 
$t \in \{0,\dots,H-1\}$ denote discrete decision steps within an episode of length $H$. 
Denote by $z_t \in \mathcal{Z}$ the global traffic state at time $t$ and by 
$\mathbf{a}_t \coloneqq (a_t^i)_{i \in \mathcal{I}}$, the joint signal action over all intersections. 
The traffic state then evolves according to the Markovian dynamics:
$
z_{t+1} \sim 
P_{\mathrm{env}}\bigl(\,\cdot \mid z_t, \mathbf{a}_t \bigr),
$
where the transition kernel $P_{\mathrm{env}}$ is induced by \textsc{SUMO}'s microscopic vehicle dynamics and the signal programs.

\textbf{Observation.}For each intersection $i \in \mathcal{I}$, the observation $o_t^i = f_i(s_t)$ summarizes traffic and signal conditions in a local neighborhood. On the traffic side, we track eight aggregated lane movements (straight/right (SR), and left-turn (LT) for North/South/East/West) and, for each movement, record eight features: a movement identifier (SR vs.\ LT), distances and speeds of the closest and second-closest vehicle to the intersection, lane density, mean speed, and a queue fraction (fraction of vehicles with $v < 0.1$\,\si{m/s}, normalized to a 10-vehicle queue). This yields $8\times 8 = 64$ traffic features. In addition, the agent obtains basic signal-state information for the controlled intersection and its four orthogonal neighbors: time since last phase change, current phase index, and a flag indicating whether the signal is in a yellow/all-red (clearance) interval, giving $5\times 3 = 15$ signal features. Together, these form a compact observation vector
\(
o_t^i \in \mathbb{R}^{79},
\)
designed to capture local pressure and limited neighborhood context.

\textbf{Action and safety.}
At each decision step, agent $i$ selects a discrete action $a_t^i \in \{0,\dots,7\}$, choosing one of eight non-conflicting signal phases at that junction. The phase set covers straight/right movements for the north--south and east--west pairs, protected left turns for these pairs, and four single-approach phases where all movements (SR+LT) on one approach receive green. Safety is enforced via a fixed \SI{5}{s} clearance interval (yellow plus all-red) applied whenever a phase changes, ensuring that vehicles clearing the intersection are never exposed to newly conflicting movements. When switching between phases, we allow ``green carryover’’ for movements that are green in both the old and new phase, avoiding unnecessary clearance for non-conflicting streams and improving throughput.

\textbf{Reward.}
The agents are homogeneous and share the same local reward specification. 
At every step $t$, each agent $i$ receives a local reward $r_t^i$ defined over vehicles on its incoming edges. 
The reward combines two components: (i) a normalized mean-speed term promoting throughput, and (ii) a penalty on normalized queue length. 
Concretely, we use
$r_t^i
= \kappa_s \, s_i
- \kappa_q \, \bar q_i,$
where $s_i$ is the normalized mean speed on incoming edges, $\bar q_i$ is a normalized queue measure, and the weights $\kappa_s,\kappa_q$ are shared across all agents.
To train the homogeneous shared policy, we use a single team reward given by the sum of local rewards,
$r_t \;=\; \sum_{i\in\mathcal{I}} r_t^i,$
which provides a common learning signal for all agents.

\subsection{Traffic demand}
\label{subsec:baseline_demand}
For initial training of the baseline policy, we use an even traffic distribution over the outer edges of the grid. Each outer edge acts as an origin, with a fixed inflow of \SI{400}{veh/h}, resulting in a total demand of \SI{4800}{veh/h}. For every vehicle spawned at origin $o$, the destination $d$ is sampled uniformly from the remaining $11$ outer edges, so that
$\Pr\{d \mid o\} = \tfrac{1}{11}, \quad d \neq o.$
Routes between each origin–destination pair follow shortest paths in edge count (breaking ties uniformly at random). This yields a balanced training distribution where all directions and origin–destination (OD) pairs are represented. This even OD pattern is the only demand configuration used during baseline training.

\subsection{PPO-based MARL controller (baseline)}
\label{subsec:ppo-baseline}

In the baseline MARL controller, we adopt a CTDE scheme with parameter sharing.
All intersections use the same stochastic traffic-signal policy 
$\pi_\theta : \mathcal{O} \to \Delta(\mathcal{A})$, where 
$\pi_\theta(a \mid o)$ denotes the probability of selecting phase 
$a \in \mathcal{A}$ given the local observation $o \in \mathcal{O}$ and shared parameters $\theta$.
At time $t$, each agent $i$ executes in a decentralized manner by sampling its phase
$a_t^i \sim \pi_\theta(\cdot \mid o_t^i)$ based solely on its local observation $o_t^i$.
The agents are homogeneous and share the same local reward.
During training, trajectories and rewards collected from all intersections are aggregated 
to perform joint updates of the shared parameters $\theta$.
This parameter-sharing CTDE design exploits the homogeneity between intersections, 
improves sample efficiency, and reduces the number of trainable parameters.

Let $\tau = (s_0,o_0^{1:|\mathcal{I}|},a_0^{1:|\mathcal{I}|},r_0,\dots,s_{H})$ denote a trajectory induced by $\pi_\theta$ and the environment dynamics, where $o_t^{1:|\mathcal{I}|} := (o_t^1,\dots,o_t^{|\mathcal{I}|})$ and similarly for $a_t^{1:|\mathcal{I}|}$. The baseline objective maximizes the expected discounted return over the training distribution of traffic scenarios:
\begin{equation}
J(\theta)
\;=\;
\mathbb{E}_{\tau \sim \pi_\theta}
\Bigg[
\sum_{t=0}^{H-1} \gamma^{t} \, r_t
\Bigg],
\label{eq:global-objective}
\end{equation}
with discount factor $\gamma\in(0,1]$. In this baseline, the demand pattern is fixed to the even distribution of Section~\ref{subsec:baseline_demand}; the controller is not yet biased toward worst-case conditions.

We optimize~\eqref{eq:global-objective} using PPO~\citep{schulman2017proximal}, a clipped policy-gradient method that is widely used in continuous-control and multi-agent settings. 
Given a mini-batch 
$B$ sampled from the replay buffer $\mathcal{D}_\theta$, 
we define the policy-gradient objective
\begin{equation}
\begin{aligned}
    J(\theta; B)
    &:= 
    \frac{1}{|B|\,|\mathcal{I}|}
    \sum_{(o_t^i, a_t^i, r_t, o_{t+1}^i)\in B}
    \sum_{i\in\mathcal{I}}
        r_t \,\log \pi_\theta(a_t^i \mid o_t^i).
\end{aligned}
\end{equation}
The policy parameters are then updated by ascending the stochastic gradient 
$\nabla_\theta J(\theta; B)$.
Coordination emerges from local observations and the use of a homogeneous, parameter-shared policy. The resulting baseline MARL controller serves as a conventional, expectation-maximizing baseline on the calibrated $3\times3$ grid. In Section~\ref{sec:method}, we build on this baseline by introducing a CB-WCE that reweights or resamples demand scenarios during training, effectively biasing learning toward worst-case conditions, while keeping the underlying PPO objective~\eqref{eq:global-objective} and update rule unchanged.


\section{Distributionally Robust MARL via Worst-Case Estimation}
\label{sec:method}

\subsection{Scenario-based robustness objective}
\label{subsec:robust-objective}

Variability in real world traffic distributions suggests that a traffic-signal controller should perform reliably across \emph{sets} of demand patterns rather than being tuned to a single nominal case. In our setting, the data-driven preprocessing in Section~\ref{subsec:baseline_demand} yields a finite set of $K=8$ representative traffic-demand scenarios,
\(
\mathcal{K} = \{0,\dots,7\},
\)
corresponding to synthetic data and a clustered pNEUMA time window with a distinct spatial flow pattern (e.g., different directional imbalances or overall loading).

For a fixed MARL policy $\pi_\theta$ and a given scenario $k \in \mathcal{K}$, we define
\[
J_k(\theta)
\;=\;
\mathbb{E}\big[ R(\tau) \,\big|\, \text{scenario } k,\; \pi_\theta \big],
\]
as the expected return when the network is operated under that scenario, where $\tau$ denotes a rollout trajectory and $R(\tau)$ its cumulative reward (a negative delay-based proxy). From these per-scenario returns we derive two evaluation metrics:
\begin{equation}
J_{\mathrm{avg}}(\theta)
\;=\;
\frac{1}{K} \sum_{k \in \mathcal{K}} J_k(\theta),
\qquad
J_{\mathrm{worst}}(\theta)
\;=\;
\min_{k \in \mathcal{K}} J_k(\theta).
\label{eq:j-avg-worst}
\end{equation}
Here $J_{\mathrm{avg}}$ captures typical performance across the eight scenarios, while $J_{\mathrm{worst}}$ captures performance under the most challenging one.

In our experiments, the baseline MARL controller is trained only under the synthetic, spatially balanced demand described in Section~\ref{sec:problem}; its $J_k(\theta)$, $J_{\mathrm{avg}}(\theta)$, and $J_{\mathrm{worst}}(\theta)$ on the traffic-demand scenarios are therefore purely \emph{evaluation} quantities, not the objective it was optimized for. As a result, the baseline may perform well near its training distribution but degrade markedly on some of the data-driven scenarios, leading to a low $J_{\mathrm{worst}}$.

Our goal in the remainder of this section is to improve $J_{\mathrm{worst}}(\theta)$ and, ideally, to also improve (or at least preserve) $J_{\mathrm{avg}}(\theta)$ by retraining the MARL policy under traffic-demand patterns that are deliberately chosen to be difficult.
To this end, we introduce a CB-WCE that adaptively reweights the traffic-demand scenarios so that training is increasingly focused on those that induce high delay. The estimator itself and its integration with PPO are described next in Sections~\ref{subsec:cb-estimator}–\ref{subsec:drmarl}.

\subsection{Contextual-bandit worst-case estimator}
\label{subsec:cb-estimator}

To improve the worst-case performance in~\eqref{eq:j-avg-worst}, we introduce a \emph{worst-case estimator} that operates on a slower timescale than the traffic-signal MARL controller. Conceptually, this agent plays an adversarial role: for a fixed signal-control policy $\pi_\theta$, it tries to select traffic-demand patterns that \emph{maximize} network-wide congestion, so that subsequent updates of $\pi_\theta$ are biased toward difficult conditions.

\textbf{Observation.}
The estimator acts once every \SI{600}{\second}, i.e., once per traffic ``window'', while the traffic-light agents act every simulation second. At the end of window $\tau$, we summarize the recent network state by an 18-dimensional feature vector
\(
s_\tau \in \mathbb{R}^{18},
\)
constructed from the average speed and average density at each of the nine intersections (two scalars per intersection). This compact summary provides coarse information about where congestion currently occurs in the grid.

\textbf{Action.}
Given the window-level observation $x_\tau$, the estimator outputs a non-negative weight vector
\[
w_\tau \in \mathbb{R}_{\ge 0}^8, \qquad \sum_{k=1}^8 w_{\tau,k} = 1.
\]
with one component $w_{\tau,k}$ for each of the $K=8$ traffic-demand scenarios. In practice, $w_\tau$ is the output of the trained neural network. These weights define a \emph{mixed} demand pattern for the next \SI{600}{\second} by combining the base inflow vectors $\{\lambda^{(k)}\}_{k=1}^8$ extracted from the inflow data:
\[
\lambda_{\mathrm{mix}}(w_\tau) \;=\; \sum_{k=1}^8 w_{\tau,k} \,\lambda^{(k)}.
\]
The resulting mixed inflows are then applied in the SUMO/Flow environment for the duration of the next window.

\textbf{Reward.}
During window $\tau{+}1$, the MARL traffic-light policy $\pi_\theta$ is applied at every second while demand is generated according to $\lambda_{\mathrm{mix}}(w_\tau)$. Over this window, we accumulate the total waiting time of all vehicles in the network,
\(
r_\tau^{\mathrm{CB}} = \sum_{v} \text{wait\_time}_v,
\)
where $\text{wait\_time}_v$ is the time vehicle $v$ spends queued in this time window  $\tau$. The CB-WCE seeks to \emph{maximize} this cumulative waiting time, in contrast to the signal controller, which aims to minimize delay.

The worst-case estimator’s policy $\pi_\psi(w \mid s)$ is parameterized by a neural network and trained with a policy-gradient method. 
Here, $\pi_\psi(w_\tau \mid s_\tau)$ denotes a probability distribution over a weight vector for different traffic-demand scenarios, conditioned on the current local observation. As shown in Algorithm~\ref{alg:worst}, training is carried out with $\theta$ frozen at $\theta_0$, using tuples $(s_\tau,w_\tau,r_\tau^{\mathrm{CB}})$ collected over successive windows. Given a mini-batch 
$B$ 
sampled from the replay buffer $\mathcal{D}_w$, we define the contextual-bandit objective
\begin{equation}
    J_{\mathrm{CB}}(\psi; B)
    :=
    \frac{1}{|B|}
    \sum_{(s_\tau, w_\tau, r_\tau^{\mathrm{CB}}, s_{\tau+1}) \in B}
    r_\tau^{\mathrm{CB}} \,\log \pi_\psi(w_\tau \mid s_\tau),
\end{equation}
and update the estimator parameters via ascending the stochastic gradient 
$\nabla_\psi J_{\mathrm{CB}}(\psi; B)$., yielding a policy $\pi_{\psi^\ast}$ that focuses on adverse traffic-demand mixtures. In the subsequent DR-MARL phase (Section~\ref{subsec:drmarl}), the worst-case estimator’s policy $\pi_{\psi^\ast}$ is kept fixed and used only to generate demand sequences.

\begin{algorithm}[t!]
\caption{Contextual-bandit worst-case estimator}
\label{alg:worst}
\begin{algorithmic}[1]

\State \textbf{Input:} OD--based group distributions $\{\lambda^{(k)}\}_{k=1}^8$;
traffic-light controller policy $\pi_\theta$;
window length $T_{\mathrm{win}}$;
learning rate $l_\psi$;
episode horizon $H_{\mathrm{W}}$;
replay buffer $\mathcal{D}_w$.

\State \textbf{Output:} Worst-case estimator policy $\pi_{\psi^\ast}(w \mid s)$.

\State Initialize policy parameters $\psi$ and buffer $\mathcal{D}_w \gets \emptyset$.

\For{episode $=1,\dots,K$}
    \State Reset SUMO/Flow network.
    \State Warm up under a random mixture of $\{\lambda^{(k)}\}$.
    \State Compute initial observation $s_0$ 

    \For{group window index $\tau = 0,\dots,$ until $\mathcal{D}_w$}
        \State Sample weights $w_\tau \sim \pi_\psi(\cdot \mid s_\tau)$.
        \State Construct group $\lambda_{\mathrm{mix}}(w_\tau) = \sum_{k=1}^8 w_{\tau,k}\lambda^{(k)}$.
        \State Apply $\lambda_{\mathrm{mix}}(w_\tau)$ for the next $T_{\mathrm{win}}$ seconds.

        \State Initialize window reward $r_\tau^{\mathrm{CB}} \gets 0$.
        \For{$t = 1,\dots,T_{\mathrm{win}}$}
            \For{each intersection $i \in \mathcal{I}$}
                \State Observe $o_t^i$ and select $a_t^i \sim \pi_{\theta_0}(\cdot \mid o_t^i)$.
            \EndFor
             \State Apply $(a_t^i)_{i\in\mathcal{I}}$ under demand $\lambda_{\mathrm{mix}}(w_\tau)$.
            \State Compute waiting time $t_{\mathrm{wait}}$ of each vehicle at $t$.
            
            \State Accumulate $r_\tau^{\mathrm{CB}} \gets r_\tau^{\mathrm{CB}} +t_{\mathrm{wait}}$.
        \EndFor

        \State Compute next estimator observation $s_{\tau+1}$.
        \State Store $(s_\tau, w_\tau, r_\tau^{\mathrm{CB}}, s_{\tau+1})$ in $\mathcal{D}_w$.

    \EndFor
    \State Sample mini-batch $B \sim \mathcal{D}_w$.
    \State Update $\psi \gets \psi + l_\psi \nabla_\psi J_{\mathrm{CB}}(\psi; B)$.
\EndFor

\end{algorithmic}
\end{algorithm}

\subsection{DR-MARL training with a frozen worst-case estimator}
\label{subsec:drmarl}

We now describe how the CB-WCE is used to obtain a DR-MARL controller. As shown in Algorithm~\ref{alg:DRMARL}, after training the baseline MARL controller $\pi_{\theta_0}$ on the synthetic, spatially balanced demand (Section~\ref{subsec:baseline_demand}) and training the contextual-bandit estimator $\pi_{\psi^\ast}(w \mid x)$ against this frozen baseline (Section~\ref{subsec:cb-estimator}), we initialize the DR-MARL controller with $\theta = \theta_0$ and perform an additional PPO fine-tuning phase in which only $\theta$ is updated. The CB-WCE $\pi_{\psi^\ast}$ is kept fixed and serves as a scheduler that selects mixed demand patterns during this retraining.

During DR-MARL fine-tuning, each training episode is partitioned into windows of length $T_{\mathrm{win}} = \SI{600}{\second}$. At the beginning of window $\tau$, we compute the observation $x_\tau$ (average speed and density at each intersection over the preceding window, or over an initial warm-up period for $\tau = 0$) and query the frozen estimator to obtain demand weights
$w_\tau \sim \pi_{\psi^\ast}(\cdot \mid x_\tau),$
which define the mixed demand pattern $\lambda_{\mathrm{mix}}(w_\tau)$ for the upcoming $T_{\mathrm{win}}$ seconds. Within this window, the MARL controller applies actions $a_t^i \sim \pi_\theta(\cdot \mid o_t^i)$ at every intersection $i$ each simulation second, while vehicles are generated according to $\lambda_{\mathrm{mix}}(w_\tau)$ and rewards $r_t$ are collected exactly as in the baseline setup. After all windows of an episode have been simulated, the resulting trajectory contributes to a PPO update of $\theta$ following the same setting in Section~\ref{subsec:ppo-baseline}. The policy parameters are then updated by ascending the stochastic gradient 
$\nabla_\theta J(\theta; B)$. Episodes are repeatedly generated in this two-timescale fashion, so that the traffic-signal policy gradually adapts to the adversarially chosen mixture of demand scenarios.

\begin{algorithm}[t!]
\caption{DR-MARL with a worst-case estimator}
\label{alg:DRMARL}
\begin{algorithmic}[1]
\State \textbf{Input:} Worst estimator $\pi_{\psi^\ast}(w \mid x)$;
baseline MARL policy parameters $\theta_0$;
OD--based group distributions $\{\lambda^{(k)}\}_{k=1}^8$
window length $T_{\mathrm{win}}$;
episode horizon $H_{\mathrm{DR}}$;
PPO learning rate $l_\theta$.

\State \textbf{Output:} Distributionally robust MARL policy $\pi_{\theta^\ast}$.

\State Initialize policy parameters $\theta \gets \theta_0$ and replay buffer $\mathcal{D}_\theta$.

\For{episode $=1,\dots,K_{\mathrm{DR}}$}

    \State Reset SUMO/Flow network.
    \State Warm up using a random mixture of $\{\lambda^{(k)}\}$.
    \State Compute initial estimator observation $x_0$.

    \For{group window index $\tau = 0,\dots,$ until $H_{\mathrm{DR}}$}

        \State Sample actions $w_\tau \sim \pi_{\psi^\ast}(\cdot \mid x_\tau)$.
        \State Construct group $\lambda_{\mathrm{mix}}(w_\tau) = \sum_{k=1}^8 w_{\tau,k}\lambda^{(k)}$.
        \State Apply $\lambda_{\mathrm{mix}}(w_\tau)$ for the next $T_{\mathrm{win}}$ seconds.

        \For{$t = 1,\dots,T_{\mathrm{win}}$}
            \For{each intersection $i \in \mathcal{I}$}
                \State Observe $o_t^i$ and sample $a_t^i \sim \pi_\theta(\cdot \mid o_t^i)$.
            \EndFor
            \State Apply $(a_t^i)_{i\in\mathcal{I}}$ under demand $\lambda_{\mathrm{mix}}(w_\tau)$.
            \State Compute $r_t$ and next observations $\{o_{t+1}^i\}_{i\in\mathcal{I}}$.
            \State Store $\big(\{o_t^i\}_{i\in\mathcal{I}}, \{a_t^i\}_{i\in\mathcal{I}}, r_t, \{o_{t+1}^i\}_{i\in\mathcal{I}}\big)$ in $\mathcal{D}_\theta$.
        \EndFor

        \State Compute next estimator observation $x_{\tau+1}$.

    \EndFor
    \State Sample mini-batch $B \sim \mathcal{D}_\theta$.
    \State Update $\theta \gets \theta + l_\theta \nabla_\theta J(\theta; B)$.

\EndFor

\end{algorithmic}
\end{algorithm}


\section{Experimental Setup: 3x3 Grid with traffic Demand}
\label{sec:setup}

\subsection{Network}
\label{subsec:network-signal}
As described in Section~\ref{sec:problem}, we use a $3\times3$ signalized grid as the testbed. The geometry is based on a contiguous subarea of central Athens covered by the pNEUMA trajectory dataset~\citep{BarmpounakisGeroliminis2020}: intersection coordinates are taken from the real network and connected by straight links so that block lengths and intersection spacing reflect typical urban values. The grid does not exactly match the true road layout (which includes one-way streets, heterogeneous lane counts, and side roads), but preserves the overall scale and orientation. The construction of traffic demand is detailed in Section~\ref{subsec:demand}.

Each bidirectional road segment is modeled with four lanes per direction to provide sufficient storage under congested conditions. The lane configuration follows a conventional urban design:
\begin{itemize}
    \item lane 0 (rightmost): straight or right turn (SR),
    \item lane 1: straight only,
    \item lanes 2--3: left turns, with lane 3 also allowing U-turns.
\end{itemize}
In the RL interface, lanes 0--1 are aggregated into a single straight/right (SR) movement and lanes 2--3 into a single left-turn (LT) movement per approach, yielding two aggregated movements (SR, LT) for each of the four approaches (N, S, E, W).

\subsection{Traffic demand}
\label{subsec:demand}

To train the worst-case estimator and the DR-MARL controller, we expose the network to a small but diverse set of OD demand patterns. All patterns are defined on the 12 outer edges of the $3{\times}3$ grid.

\textbf{pNEUMA-based pattern.}
We first construct a single data-driven OD pattern from the pNEUMA trajectory dataset for central Athens~\citep{BarmpounakisGeroliminis2020}. Vehicle trajectories recorded at \SI{24}{fps} are mapped to the nine-intersection study area, and for each recording window we count the number of trips between every origin–destination pair. Aggregating the 20 windows and converting counts to hourly rates yields an OD-rate vector that reflects typical flows in the subnetwork. This vector is then normalized to a total demand of \SI{5000}{veh/h} and used as the pNEUMA-based scenario.

\textbf{Synthetic patterns.}
The aggregated pNEUMA pattern is relatively balanced and does not cover strongly directional or highly congested cases by itself. To probe robustness under a wider range of conditions, we therefore design seven additional synthetic OD patterns. Each pattern is a $12\times12$ OD matrix with the same total inflow of \SI{5000}{veh/h}, but with different spatial structure: one nearly uniform pattern, inbound and outbound patterns concentrating flow toward or away from the central area, north–south and east–west corridor patterns, and two diagonal “cross-town’’ patterns. Small random perturbations are added to avoid perfectly regular grids while keeping flows non-negative. 

\textbf{Scenario set.}
Together, the one pNEUMA-based pattern and the seven synthetic patterns form eight base demand scenarios. This set constitutes the scenario index $\mathcal{K}$ used in Section~\ref{sec:method}: during worst-case estimator training and DR-MARL fine-tuning, the estimator selects or mixes these scenarios to generate the traffic demand applied in simulation.


\subsection{Training protocol}
\label{subsec:training-protocol}

For the baseline MARL controller, we use an episode horizon of $H = 900$\,s, matching the pNEUMA window length. Each training iteration collects $N_{\text{roll}} = 10$ rollouts, yielding a total simulated time of $H \times N_{\text{roll}} = 9000$\,s per iteration. Training is ran for $3000$ iterations. The policy is initialized randomly and learns from scratch, without any pre-training or imitation.

For the CB-WCE, we use longer episodes that span multiple \SI{600}{\second} windows so that the estimator can observe the effect of its scenario-selection decisions. An episode horizon of $H_{\text{WCE}} = 9600$\,s is used, with $N_{\text{roll}}^{\text{WCE}} = 8$ rollouts per training iteration and a total of $50$ training iterations. In all cases, the per-window duration remains \SI{600}{\second}, and the eight traffic-demand scenarios are switched according to the estimator’s weights as described in Section~\ref{subsec:demand}.

For the final DR-MARL model, we adopt the same multi-window structure as for the worst-case estimator so that the policy is trained under demand that can change every \SI{600}{\second}. An episode horizon of $H_{DR}= 9600$\,s is used, with $N_{\text{roll}}^{\text{DR}} = 2$ rollouts per training iteration and a total of $400$ training iterations. The DR-MARL policy is initialized from the trained baseline MARL controller and further updated under the adaptive scenario selection described in Section~\ref{subsec:drmarl}.

\subsection{Baselines and Evaluation Metrics}
\label{subsec:baseline_eval}
In our experiments we compare two controllers:
\begin{itemize}
    \item \textbf{PPO MARL:} the expectation-maximizing multi-agent PPO controller described in Section~\ref{sec:problem}.
    \item \textbf{DR-MARL:} the same PPO architecture retrained with the CB-WCE, using the eight traffic-demand scenarios.
\end{itemize}

We evaluate performance using standard network-level traffic metrics:
\begin{itemize}
    \item \textbf{Queue length:} total number of queued vehicles.
    \item \textbf{Average speed:} mean vehicle speed over all vehicles.
\end{itemize}
To evaluate performance over the eight traffic-demand groups, we compute the horizon-average of each metric for every group and report the per-group improvement of DR-MARL over the baseline. The corresponding evaluation results will be presented in the following section.

\section{Results and Discussion}

\subsection{Results}
\label{subsec:results}
We compare the baseline MARL controller and the DR-MARL controller on the seven synthetic traffic-demand groups introduced in Section~\ref{subsec:robust-objective} (groups 0-6), the pNEUMA traffic demand introduction in Section~\ref{subsec:robust-objective} (group 7), and on one additional group derived from the enriched Sioux Falls scenario~\citep{chakirov2014enriched} (group 8). For the latter, we consider a subnetwork in Sioux Falls that is consistent with our $3\times3 $ grid and normalize the total demand to \SI{5000}{veh/h}. The DR-MARL controller is trained on the distribution groups defined in Section~\ref{subsec:demand}, whereas the baseline MARL controller is trained on the distribution defined in Section~\ref{subsec:baseline_demand}; the Sioux Falls-based group is used exclusively for evaluation as an independent, previously unseen demand pattern for both controllers. For each controller and each group $k\in\{0,\dots,8\}$, we run $10$ independent evaluation rollouts of length $H_{\mathrm{eval}} = 3600$\,s. During a rollout, the demand is fixed to group $k$ for the full horizon, the learned policy is kept fixed, and only the stochasticity in vehicle generation and microscopic dynamics varies across rollouts. For every group and controller, we then compute the horizon-average of the two metrics in Section~\ref{subsec:baseline_eval}: average queue length and average speed, averaged again over the $10$ rollouts.

\begin{figure}[t!]
    \centering
    \includegraphics[width=1.0\linewidth]{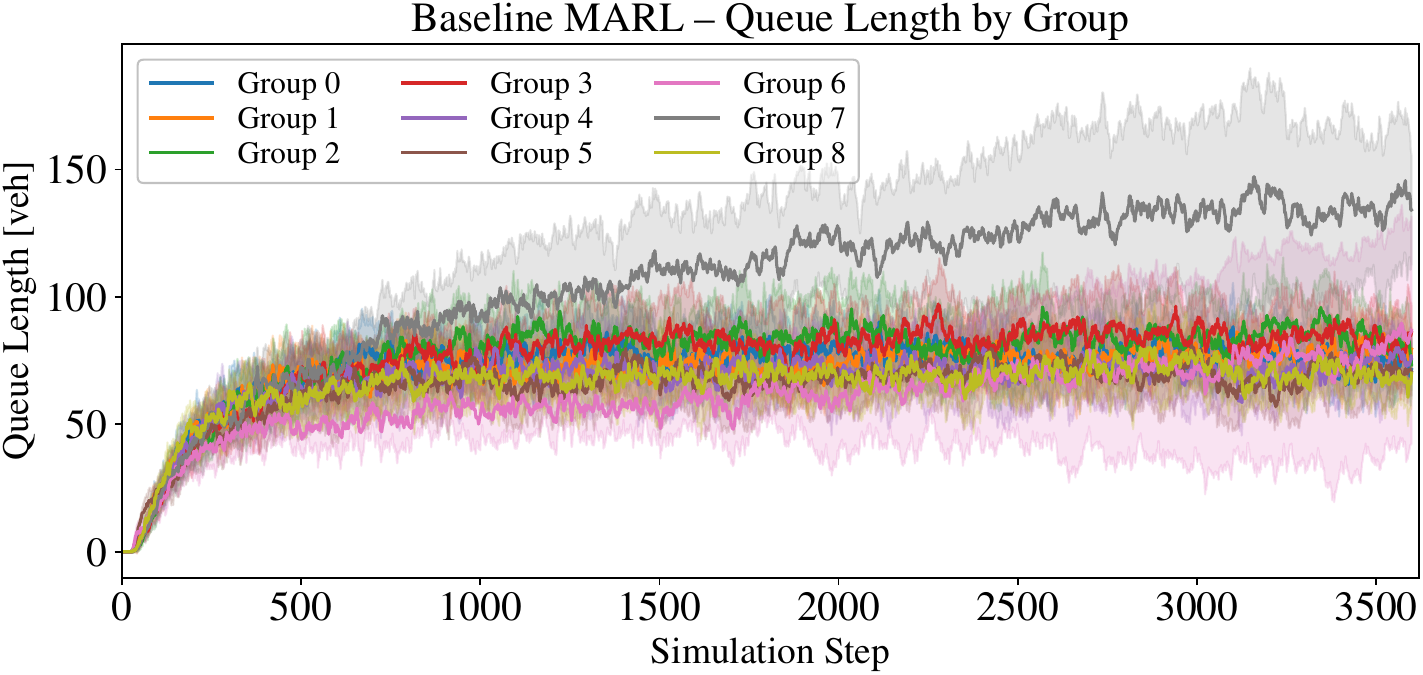}
    \caption{Network-wide queue length over time for each demand group under the baseline MARL controller.}
    \label{fig:queue_baseline}
\end{figure}

\begin{figure}[t!]
    \centering
    \includegraphics[width=1.0\linewidth]{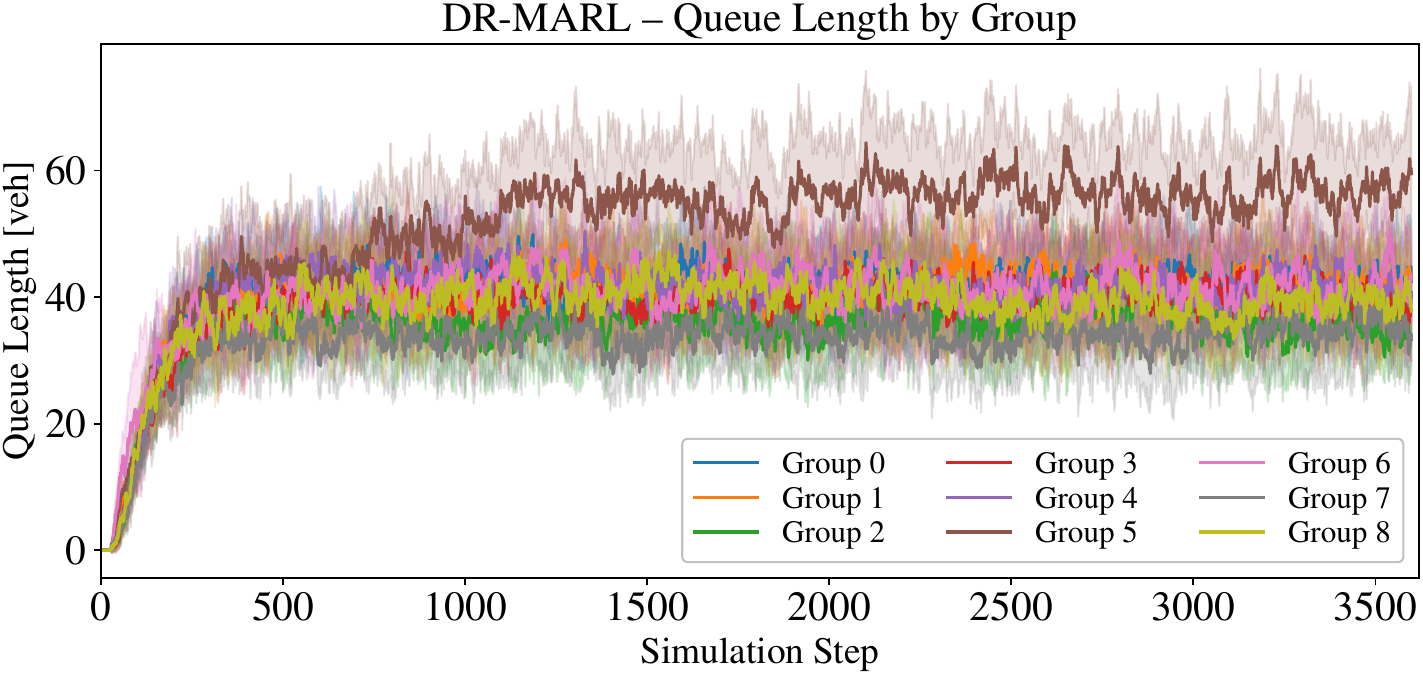}
    \caption{Network-wide queue length over time for each demand group under the DR-MARL controller.}
    \label{fig:queue_robust}
\end{figure}

Figures~\ref{fig:queue_baseline}–\ref{fig:queue_robust} show the evolution of network-wide queue length over time for all nine groups under the baseline MARL controller and the DR-MARL controller. Each coloured line corresponds to the per-timestep mean queue length (over 10 rollouts) for a single demand group. Under the baseline MARL controller, queues diverge substantially across groups and reach the highest levels for group~7, with several other groups also sustaining long queues. Under DR-MARL, the curves are consistently shifted downward: all groups stabilize at noticeably lower queue levels, with group~5 being the heaviest but still far below the baseline MARL controllers worst case. The unseen Sioux Falls group (group~8) behaves similarly to the training groups and also exhibits a clear reduction in queues under DR-MARL.

\begin{figure}
    \centering
    \includegraphics[width=1.0\linewidth]{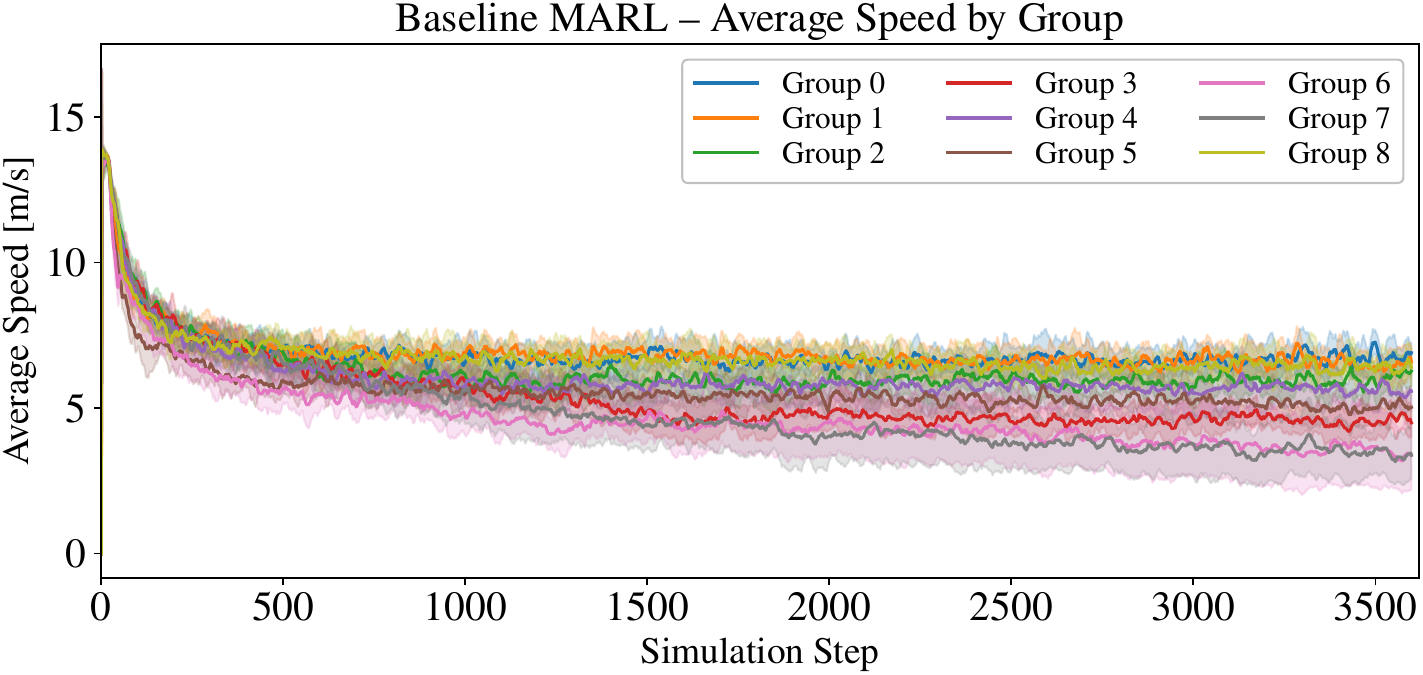}
    \caption{Network-wide average speed over time for each demand group under the baseline MARL controller.}
    \label{fig:speed_baseline}
\end{figure}

\begin{figure}
    \centering
    \includegraphics[width=1.0\linewidth]{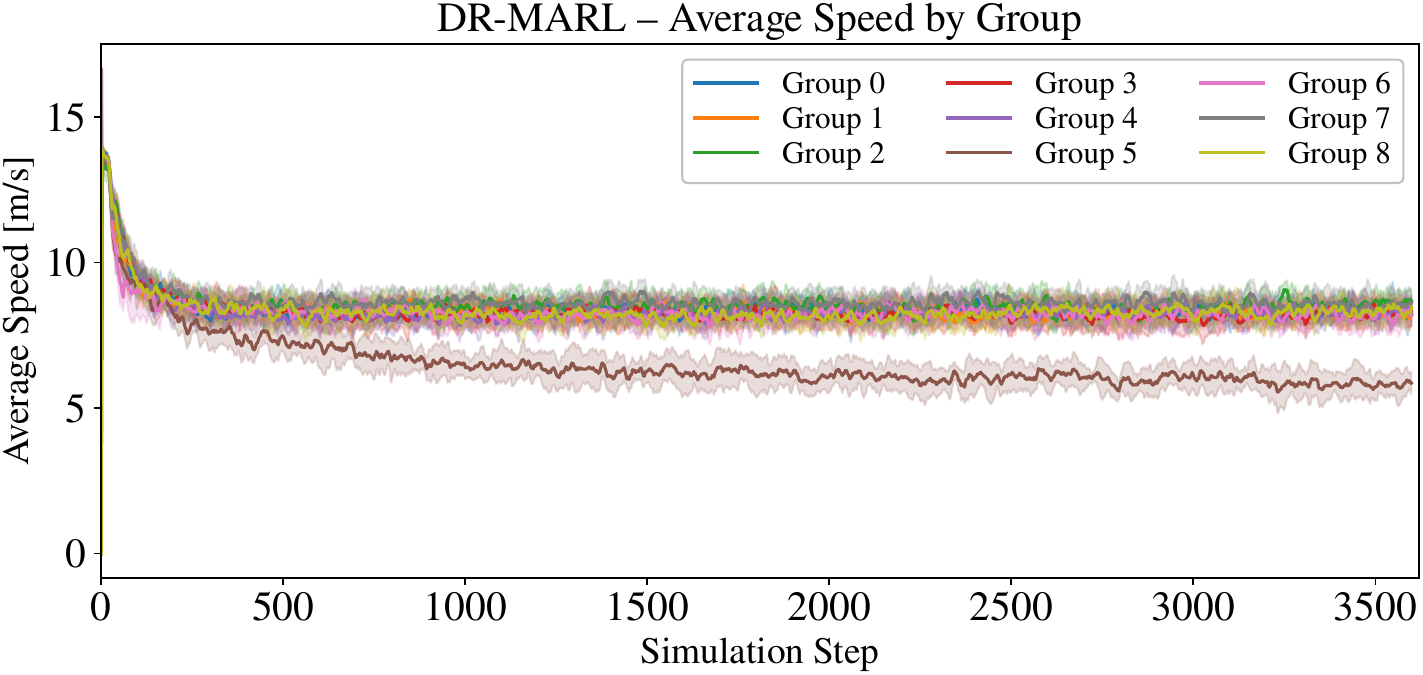}
    \caption{Network-wide average speed over time for each demand group under the DR-MARL controller.}
    \label{fig:speed_robust}
\end{figure}

Figures~\ref{fig:speed_baseline} and~\ref{fig:speed_robust} report the corresponding average speeds. Under the baseline MARL controller, steady-state speeds vary between roughly \SIrange{4.5}{7}{m/s} depending on the group, with groups~6 and~7 performing worst. Under DR-MARL, all nine groups attain higher steady-state speeds, typically around \SI{8}{m/s} or above, with only group~5 clearly below this level. The Sioux Falls group again follows this pattern, confirming that the DR-MARL policy generalises better than the baseline to an unseen demand distribution.

Table~\ref{tab:comp} summarizes these results numerically by reporting, for each group, the horizon- and rollout-averaged queue length and average speed under both controllers. The DR-MARL controller substantially reduces the average queue length for all groups and at the same time increases the average speed in every group.

\begin{table}[t!]
    \centering
    \caption{Comparison of MARL and DR-MARL performance per group (horizon-and-rollout-averaged).}
    \label{tab:comp}
    \begin{tabular}{|c|cc|cc|}
        \hline
        & \multicolumn{2}{c|}{queue length} & \multicolumn{2}{c|}{average speed} \\
        \cline{2-5}
        groups\rule{0pt}{2.6ex}
        & MARL & DR-MARL & MARL & DR-MARL \\
        \hline
        0 & 72.287 & 40.137 & 6.882 & 8.411 \\
        1 & 70.318 & 39.959 & 6.925 & 8.404 \\
        2 & 76.829 & 35.250 & 6.297 & 8.571 \\
        3 & 76.062 & 38.528 & 5.429 & 8.358 \\
        4 & 67.008 & 40.250 & 6.105 & 8.347 \\
        5 & 64.620 & 51.302 & 5.647 & 6.524 \\
        6 & 60.094 & 39.914 & 4.714 & 8.330 \\
        7 & 105.153 & 32.932 & 4.910 & 8.685 \\ %
        8 & 65.711 & 38.372 & 6.795 & 8.353 \\
        \hline
    \end{tabular}
\end{table}

Table~\ref{tab:comp_diff} reports the relative change of DR-MARL with respect to the baseline (percentage decrease in queue length and percentage increase in average speed). Queue length decreases by roughly $21$–$69\,\%$ across groups, with the largest reduction in group~7 ($-68.68\,\%$). Average speed increases by roughly $16$–$77\,\%$, with the largest gains in groups~6 and~7 (about $+76.7\,\%$). For the unseen Sioux Falls group (group~8), DR-MARL reduces queues by about $41.6\,\%$ and increases speed by about $22.9\,\%$.

\begin{table}[t!]
    \centering
    \caption{Relative decrease in queue length and increase in average speed of DR-MARL compared to MARL (in~\%).}
    \label{tab:comp_diff}
    \begin{tabular}{|c|c|c|}
        \hline
        & queue length & average speed \\
        \cline{2-3}
        groups\rule{0pt}{2.6ex}
        & decrease & increase \\
        \hline
        0 & -44.475 & 22.208 \\
        1 & -43.173 & 21.344 \\
        2 & -54.119 & 36.104 \\
        3 & -49.346 & 53.956 \\
        4 & -39.933 & 36.722 \\
        5 & -20.611 & 15.536 \\
        6 & -33.581 & 76.690 \\
        7 & -68.682 & 76.693 \\
        8 & -41.605 & 22.931 \\
        \hline
    \end{tabular}
\end{table}

\begin{figure}[t!]
    \centering
    \includegraphics[width=1.0\linewidth]{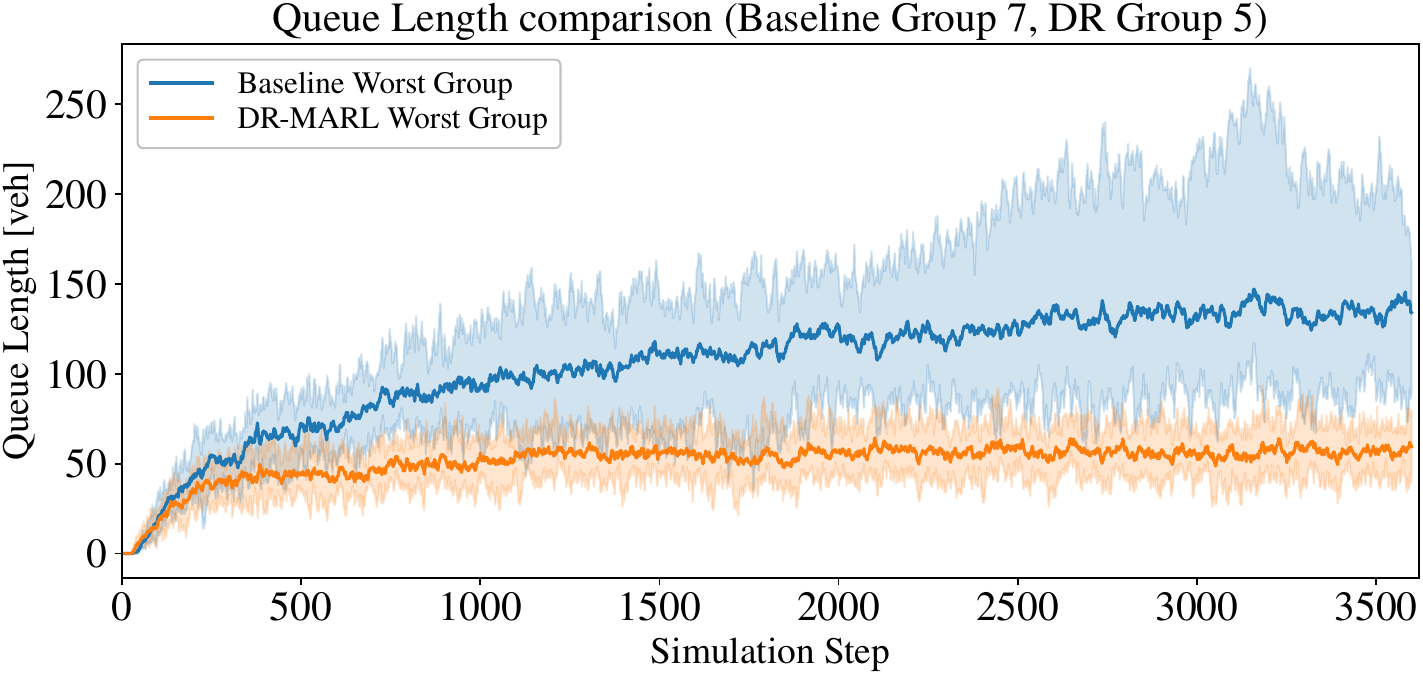}
    \caption{Queue length over time for the worst-performing baseline group and the worst-performing DR-MARL group.}
    \label{fig:queue_worst}
\end{figure}

\begin{figure}[t!]
    \centering
    \includegraphics[width=1.0\linewidth]{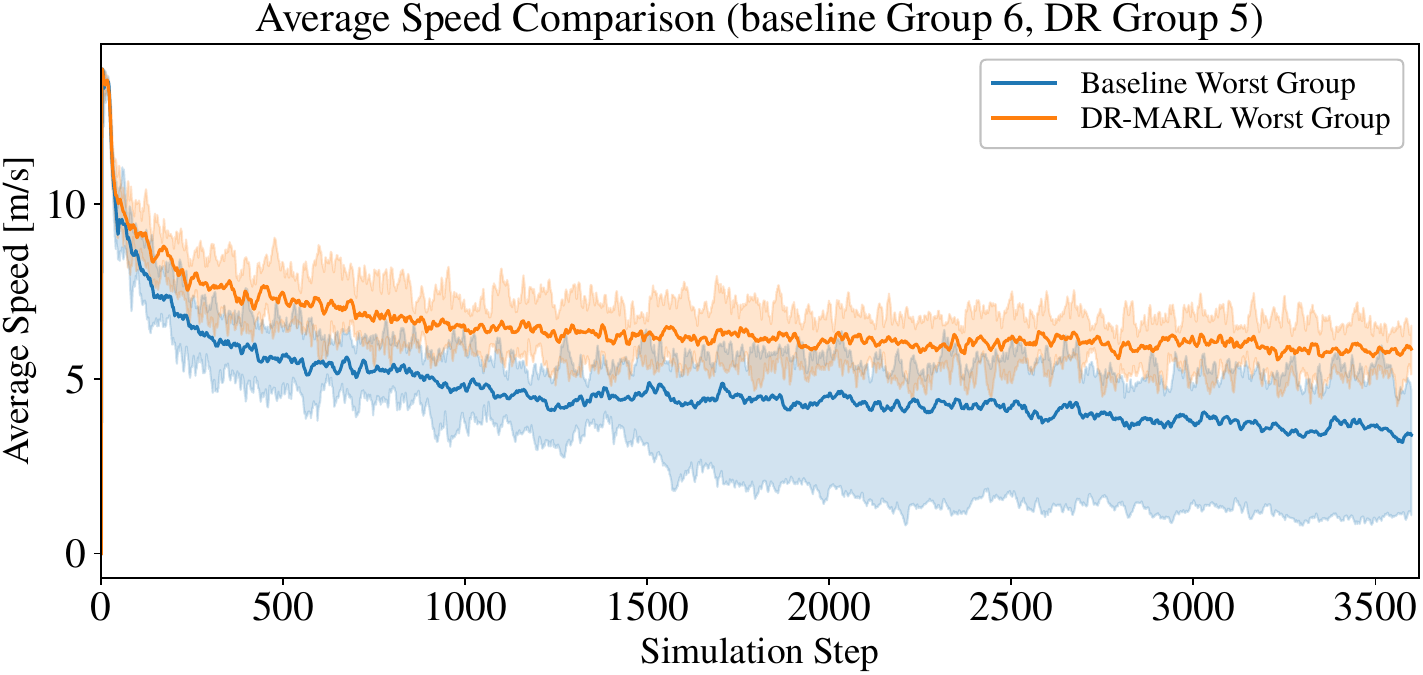}
    \caption{Average speed over time for the worst-performing baseline group and the worst-performing DR-MARL group.}
    \label{fig:speed_worst}
\end{figure}

As discussed in Section~\ref{subsec:robust-objective}, our main interest lies in improving the worst-case performance $J_{\mathrm{worst}}$ across demand groups. From Table~\ref{tab:comp}, we identify group~7 as the worst-performing group in terms of average queue length and group~6 as the worst for average speed under the baseline MARL controller, whereas group~5 is the worst group for both metrics under the DR-MARL controller. Comparing these worst cases, the DR-MARL controller reduces the worst observed average queue length by about $51.2\,\%$ (baseline group~7 vs.\ DR-MARL group~5) and increases the worst observed average speed by about $38.4\,\%$ (baseline group~6 vs.\ DR-MARL group~5). Figures~\ref{fig:queue_worst} and~\ref{fig:speed_worst} visualize the evolution of queues and speeds over time for these worst-case baseline and DR-MARL groups. The DR-MARL trajectories maintain substantially lower queues and higher speeds over most of the horizon, indicating that the contextual-bandit reweighting improves the policy precisely on its most challenging scenarios.

\subsection{Discussion}
\label{subsec:discussion}

The results show that distributionally robust retraining with the contextual-bandit estimator improves both typical and worst-case performance: DR-MARL reduces queues and increases speeds across all nine demand groups, including the unseen Sioux Falls pattern, with large gains on the baseline’s most challenging groups. These improvements are obtained without modifying the MARL architecture or reward, but solely by changing which demand patterns the policy encounters during fine-tuning. At the same time, Figures~\ref{fig:queue_worst} and~\ref{fig:speed_worst} show that the shaded regions for the two controllers overlap slightly, meaning that while DR-MARL is better for most rollouts and most of the horizon, some individual rollouts perform similarly to, or marginally worse than, the baseline MARL controller. This residual overlap is expected given the stochastic nature of the simulations and the finite set of demand groups used during training.

\section{Conclusion and Outlook}
\label{sec:conclusion}

We studied distributionally robust multi-agent reinforcement learning for traffic-signal control on a $3\times3$ grid calibrated from the pNEUMA dataset. Building on a baseline MARL controller, we introduced a CB-WCE that runs on a slower time scale, observing aggregate speed and density and selecting mixture weights over eight demand scenarios. These mixtures are used during fine-tuning, improving robustness without changing the MARL architecture or reward in the SUMO/Flow environment.

Numerical experiments on nine distinct traffic-demand patterns show that the DR-MARL controller improves both typical and worst-case performance compared to the baseline. Across nine demand patterns (including an unseen Sioux Falls-based distribution), DR-MARL reduces horizon-averaged queues and increases average speeds, with group-wise improvements of about $21$–$69,\%$ in queue length and $16$–$77,\%$ in speed, corresponding to a higher $J_{\mathrm{avg}}$ and a clear improvement in $J_{\mathrm{worst}}$. Some rollout-level overlap between the two controllers remains, but the overall distribution of outcomes shifts markedly in a favourable direction.


The present study has several limitations. Robustness is assessed on a finite set of hand-crafted demand scenarios on a stylized \(3\times 3\) grid, which restricts the diversity of operating conditions and network structures that are represented. In addition, the worst-case estimator is trained only against the baseline MARL controller and then kept fixed during DR-MARL fine-tuning, so the demand patterns it selects reflect worst cases of the baseline rather than those of the improved controller. Extending the framework to larger and more heterogeneous networks, and allowing the worst-case estimator to adapt to the evolving DR-MARL policy, is left for future work.

\bibliography{ifacconf}      

\end{document}